\documentclass[12pt,preprint]{aastex}
\usepackage{lscape}

\begin{document}
\title{Post-Decadal White Paper: A Dual-Satellite Dark-Energy/Microlensing 
NASA-ESA Mission}

\author{Andrew Gould}

\affil{
Dept.\ of Astronomy, Ohio State University, 
140 West 18th Avenue, Columbus, OH 43210\\
gould@astronomy.ohio-state.edu
}

\begin{abstract}

A confluence of scientific, financial, and political factors
imply that launching two simpler, more narrowly defined 
dark-energy/microlensing satellites will lead to faster, cheaper, better
(and more secure) science than the present
EUCLID and WFIRST designs.  The two satellites, one led by
ESA and the other by NASA, would be explicitly designed to
perform complementary functions of a single, dual-satellite
dark-energy/microlensing ``mission''.  One would be a purely optical
wide-field camera, with large format and small pixels, optimized
for weak-lensing, which because of its simple design, could
be launched by ESA on relatively short timescales. The second
would be a purely infrared satellite with marginally-sampled 
or under-sampled pixels,
launched by NASA.  Because of budget constraints, this would
be launched several years later.  The two would complement
one another in 3 dark energy experiments (weak lensing, baryon
oscillations, supernovae) and also in microlensing planet
searches.  Signed international agreements would guarantee
the later NASA launch, and on this basis equal access of both
US and European scientists to both data sets.

\end{abstract}

\section{Introduction: Faster, Cheaper, ``Better''}

In an earlier pre-Decadal-Survey white paper \citep{gould09}, I advocated
a dark energy (DE) space mission that would perform only one
DE experiment, as opposed to a ``mega-mission''
trying to simultaneously perform 3 experiments (with fairly incommensurate
optimal design specifications).  Regarding DE
per se, I expressed indifference as to whether this satellite
should carry out a weak lensing (WL), baryon acoustic oscillation (BAO),
or supernova (SN) experiment.  I noted that all 3 have advantages and
disadvantages.  However, I pointed out that satellites designed for
WL were actually optimal for also doing microlensing ($\mu$L)
planet searches, and even the times of year when WL and $\mu$L targets
are available were noted to be complementary.

The basic point was that a simple satellite would be launched sooner
and, by providing earlier results, stimulate both theoretical
ideas on DE and broader interest in (and so broader support
for) new missions that could carry out additional DE
experiments.  By contrast, the more complex satellite designs that
seemed to be taking shape on both sides of the Atlantic would
lead to technical delays that would inevitably be compounded by the 
adverse financial environment.   Analogy was made to WMAP, which
was clearly both ``faster and cheaper'' and, as a result,
``better'' than a more complex mission because it produced earlier
results that stimulated interest in later, more complex missions.

The recommendations of the Decadal Committee \citep{blandford10}
were in basic accord with the viewpoint advanced in my white paper.
The WFIRST design is far simpler than the do-everything 
(and so, ultimately, do-nothing) mega-missions that were emerging.
It is true that the WFIRST design still envisages 3 DE experiments, but
it achieves this by somewhat restricting WFIRST's capabilities
in favor of planned synergies with ground-based observations.

However, the purely space-based requirements on WFIRST are still somewhat
contradictory.  Most notably, the pixel scale must be a compromise
between the high-resolution needed for WL and the large angular area
needed for BAO and SN.  $\mu$L is the main beneficiary of this
compromise, since its pixel-scale requirements are intermediate.
Moreover, as I will summarize in the next section, WFIRST does
not fit into the immediate financial priorities of the US,
which substantially diminishes its independent role in pushing
forward DE science and so, indirectly, diminishes the chance that
it will ever be launched.

\section{New Requirements: Faster, Cheaper, ``Better''}

\subsection{NASA Response to Astro2010: Germ of an Excellent Approach}

NASA's basic response to Astro2010 has been
\hfil\break\noindent{1)} Great idea!
\hfil\break\noindent{2)} Budget does not permit immediate start on 
WFIRST (\$1.5B)
\hfil\break\noindent{3)} Can implement Decadal Science by
\hfil\break\indent{A)} Modest (20--33\%) share in EUCLID
\hfil\break\indent{B)} Begin work on WFIRST after JWST is launched

This is a reasonable response given NASA's fiscal constraints, but
in its current form, makes no sense scientifically.  For example, if 
EUCLID is launched in its current mega-mission form, the scientific
drivers for launching WFIRST 5 years later will be muted at best.
And this muting, combined with financial pressures and other
competing scientific agendas, means that WFIRST would in fact
never be launched.

Nevertheless, this response does contain the germ of an approach 
that can yield outstanding science, while taking account of the
financial and political realities.  As in the NASA response, there
would be two satellites, but these would be explicitly designed to
be complementary (constituting a single ``mission'')
and would be linked by binding international
agreements.  The first satellite, undertaken by ESA, would
be a purely optical wide-field version of EUCLID, optimized  for WL,
i.e., with excellent resolution.  The second satellite, undertaken
by NASA, would be a purely infrared telescope, basically in the model
of WFIRST but with weaker constraints on image resolution (and so with
larger field of view).  In the following subsections, I outline
how this ``dual-satellite mission'' is nearly optimal from a science 
standpoint, while meeting all the financial and political constraints.

\subsection{Science}

There are four major scientific objectives, WL, BAO, SN (for DE) and
$\mu$L (for planets).  I treat these in turn.

WL would benefit the most.  If the ESA mission were simplified to
a wide-field optical imager, it would be substantially cheaper and
easier to build and so would be launched sooner.  Freed from the
burden of infrared (and other) add-ons, it could probably have
a larger field or a larger mirror or a longer operational lifetime
(or all three), but in any case would
be launched faster and at lower risk.  There would be immediate
science return.  It is true that the science would not be as
strong as it would be with complementary infrared (IR) 
photometry for photometric
redshifts, but these would eventually be forthcoming when WFIRST
was launched.  So there would be initial excellent returns and
improvements after several years.  The second stage of analysis
based on IR data is not likely to be quickened (and may be further
delayed) by having a single EUCLID mega-mission because it would
be more complex, more expensive, and more subject to delays or even
cancellation.  Hence, there are substantial gains, reduced risk,
and very little lost with this approach.

BAO and SN would also benefit.  The basic point here is that the
WFIRST-like mission in this scenario is launched on the same
timescale as within the NASA response.  Its IR capability
would be crucial to BAO photo-$z$'s for $1.5<z<2.5$ and
for rest-frame-IR SN lightcurves at $z<0.8$ (in addition to WL
photo-$z$'s mentioned above).  WFIRST's ability to meet
BAO and SN objectives, by coordinating with ground-based observations 
is exactly the same as within the Decadal Committee's recommendation.
In fact there are two improvements: first the field of view can be
somewhat larger because the IR imager pixel size is not constrained
by WL high-resolution requirements.  Second, there will be high-resolution
optical images of essentially all fields, which could help the analysis
in some aspects.  Again, if one is extremely optimistic about how
fast a mega-mission EUCLID can be launched, then there will be some
delay in achieving these objectives.  But if NASA is bound by international
agreements to build and launch WFIRST, these delays will be modest
at worst compared to the delays likely from trying to build a mega-mission
EUCLID.

$\mu$L has significant benefits but some costs.  By streamlining
EUCLID, a $\mu$L planet search gets on the sky much faster than it
would otherwise.  For fixed aperture, field of view, and duration, an optical
$\mu$L planet search will monitor  perhaps 2--3 times fewer microlensing
events.  However, first, the full power of the IR survey will follow
when WFIRST is launched. Second, it is likely that the optical
camera on EUCLID can actually be bigger than the IR camera on WFIRST.
Third, the higher resolution in the optical, while not essential
to microlensing, will be of some benefit.  Fourth, detection of
host-star light, which is the main way that host-star masses will
be estimated, is easier if there is a longer time baseline, which
will be a byproduct of having time-separated satellites looking at
the same field.  

There are, however, several drawbacks for $\mu$L.  First, once
the WL constraints on IR pixel size are relaxed, then $\mu$L no
longer benefits from the ``compromise'' between high resolution and
wide field. $\mu$L can tolerate marginally sampled (or even slightly
under-sampled) images because the source shapes are known a priori
($\delta$-functions convolved with the PSF).  But if the image quality
is degraded too far, $\mu$L light curves will suffer.  Further study
will be required to determine if there are real conflicts between 
the BAO/SN and $\mu$L requirements on IR pixel size.  If there are,
some compromise will be in order.  Second, the field choice of
an IR-only or optical-only $\mu$L experiment would not be the same:
IR can probe much closer to the densest star fields toward the
Galactic center, which have a higher $\mu$L rate.  Therefore, either
some compromise fields would have to be chosen, or the wider optical
telescope could concentrate on an area of lower reddening while
doing ``supplementary'' observations of the designated IR fields.
This is a complication, but not an insurmountable one.

\subsection{Finance}

Clearly the best financial aid that NASA can give ESA for EUCLID
is not to supply IR arrays, but to relieve ESA of the requirement
of incorporating them in the mission.  This saves not only the
fabrication costs, but also reduces the complexity of optics,
spacecraft design, weight, communications, etc.  The main
financial concern on the NASA side is not absolute cost
but timing.  By committing itself to build WFIRST on a specified
schedule, NASA rids itself of the 20\% (or 33\%) share of EUCLID,
and wins equal access to EUCLID data for US scientists.

Better still would be $\sim 20\%$ NASA-ESA cross contributions
for the two satellites.  This would provide the basis for
science cross participation, while retaining unitary management
for each satellite.  Moreover, the broad specifications of the
two satellites should be jointly worked out on the basis of
optimal science.  If these lead to substantially different
estimated costs, the cross contributions could be
adjusted to make the financial burdens truly equal.

\subsection{Politics}

My original proposal of
first launching a simpler satellite to carry out a single DE
experiment has apparently been a non-starter on both
sides of the Atlantic, for fundamentally political reasons:
with the high cost and infrequent launching of major missions,
it is natural to fear that ones ``own'' approach to DE will
never get a mission unless it is on the first such mission.
Hence, any single-experiment satellite is guaranteed to acquire
more political enemies than friends.  However, if there is
a signed international agreement to launch a follow-up satellite,
and if it is explicitly recognized by all parties that the
two satellites constitute a single inseparable ``mission'' 
(in the larger sense), then this concern goes away, or at least
is strongly diminished.

\section{Conclusion}

Dark energy is arguably the most important physics problem of the 
21st century, with major implications for astronomy, fundamental physics, 
and perhaps even philosophy.  Significant progress will almost certainly
require unprecedented levels of international cooperation among 
scientists and the agencies that both fund and provide an organizational
framework for their largest undertakings.  Cooperation is required not
only to mount experiments of sufficient scope, but also to rationally
divide work into complementary components that enable a multi-faceted
attack while minimizing the complexity of individual facilities.

Because the required level of cooperation is unprecedented, it will not be
achieved easily.  The structures and procedures of ESA and NASA
do not lend themselves to coordinated missions.  Nor is it
the habit of astronomers to think in these terms.  But the stakes are
high.  Without bold leadership that can draw on diverse efforts to
build a coordinated attack on the problem, dark energy studies are
likely to sputter for years or decades.

\acknowledgments I thank David Weinberg for helpful discussions.





\begin{thebibliography}{99}

\bibitem[Blandford et al.(2010)]{blandford10} Blandford, R. et al. 2010,
``New Worlds, New Horizons in Astronomy and Astrophysics'', National
Academies Press

\bibitem[Gould(2009)]{gould09}  Gould, A. 2009, 
``Wide Field Imager in Space for Dark Energy and Planets'', 
2009astro2010S.100G, arXiV:0902.2211

\end{thebibliography}
\end{document}